\UseRawInputEncoding
\documentclass[aps,prl,showpacs,twocolumn]{revtex4-1}
\usepackage{graphicx}
\usepackage{dcolumn}
\usepackage{bm}
\usepackage{color}
\usepackage{amsmath}
\usepackage{amssymb}

\begin{document}
\title{Spectrally resolved  NOON state interference }
\author{Rui-Bo Jin$^{1,2}$, Ryosuke Shimizu$^{3}$,  Takafumi Ono$^{2,4,5}$,   Mikio Fujiwara$^{2}$,  Guang-Wei Deng$^{6}$}
\author{Qiang Zhou$^{6}$ }
\email{zhouqiang@uestc.edu.cn}
\author{Masahide Sasaki$^{2}$ }
\email{psasaki@nict.go.jp}
\author{Masahiro Takeoka$^{2,7}$}
\email{Takeoka@uec.ac.jp}

\affiliation{$^{1}$Hubei Key Laboratory of Optical Information and  Pattern Recognition, Wuhan Institute of Technology, Wuhan 430205, China}
\affiliation{$^{2}$National Institute of Information and Communications Technology, 4-2-1 Nukui-Kitamachi, Koganei, Tokyo 184-8795, Japan}
\affiliation{$^{3}$University of Electro-Communications, 1-5-1 Chofugaoka, Chofu, Tokyo 182-8585, Japan}
\affiliation{$^{4}$Faculty of Engineering and Design, Kagawa University, 2217-20, Hayashi-cho, Takamatsu, Kagawa Japan}
\affiliation{$^{5}$JST, PRESTO, 4-1-8 Honcho, Kawaguchi, Saitama, 332-0012, Japan}
\affiliation{$^{6}$Institute of Fundamental and Frontier Sciences and School of Optoelectronic Science and Engineering,\\ University of Electronic Science and Technology of China, Chengdu 610054, China}
\affiliation{$^{7}$Keio University, 3-14-1 Hiyoshi, Kohoku, Yokohama, Kanagawa 223-8522, Japan}

\date{\today }

\begin{abstract}
NOON state interference (NOON-SI) is a powerful tool to improve the phase sensing precision, and plays an important role in quantum measurement.
In most of the previous NOON-SI experiments, the measurements were performed in time domain where the spectral information of the involved photons was integrated and lost during the measurement.
In this work, we experimentally measured the joint spectral intensities (JSIs) at different positions of the interference patterns in both time and frequency domains.
It was observed that the JSIs were phase-dependent and show  odd (even)-number  patterns at $0$ ($\pi$) phase shift;
while no interference appeared in time domain measurement,
the interference pattern clearly appeared in frequency domain.
To our best knowledge, the latter is the first observation of the spectrally  resolved NOON state interference, which provides alternative information that cannot be extracted from the time-domain measurement.
To explore its potential applications, we considered the interferometric sensing with our setup.
From the Fisher information-based analysis, we show that the spectrally resolved NOON-SI has a better performance at non-zero-delay position than its non-spectrally resolved counterpart.
The spectrally resolved  NOON-SI scheme may be useful for quantum metrology applications such as quantum phase sensing, quantum spectroscopy, and remote synchronization.
\end{abstract}

\maketitle

\textbf{\emph{Introduction}}
\noindent
Multi-photon entanglement and multi-photon interference are useful nonclassical resources in quantum information applications.
In particular, the so-called NOON state interference (NOON-SI) is a powerful tool to improve the phase sensing precision.
NOON state is an entangled state with  $N$ photons occupying either one of two optical modes (e.g., polarization modes
 or path modes): $\frac{1}{{\sqrt 2 }}(\left| {N0} \right\rangle  + \left| {0N} \right\rangle )$  \cite{Boto2000, Edamatsu2002, Giovannetti2002}.
NOON-SI is a powerful tool to improve  the phase sensing precision to Heisenberg limit of $\Delta \phi=1/N$ , much higher than the shot noise limit of $\Delta \phi=1/\sqrt{N}$, which is precision limit of  the classical interferometer, e.g., the  Mach-Zehnder interferometer (MZI) \cite{Giovannetti2004}.
NOON-SI has been widely used in  super-resolving phase
measurements \cite{Mitchell2004, Walther2004, Nagata2007, Afek2010, Slussarenko2017, Zhou2017PRAppl}, quantum lithography \cite{Boto2000}, quantum microscopy \cite{Ono2013, Israel2014},  quantum spectroscopy \cite{Jin2018Optica}  and  error correction \cite{Bergmann2016}.

Most of the previous NOON-SI experiments were measured in time domain \cite{Boto2000, Edamatsu2002, Giovannetti2002, Giovannetti2004,
Mitchell2004, Walther2004, Nagata2007, Afek2010, Slussarenko2017, Zhou2017PRAppl, Ono2013, Israel2014, Jin2018Optica, Bergmann2016}, i.e., the interference patterns are obtained by recording the  coincidence counts as a function of the temporal delay.
However, the conjugate parameter of time, i.e., the spectral information of the involved photons was integrated and lost during the time-domain measurement.
The knowledge of the spectral correlations of the interfering photons can reveal some important information about the interference visibility \cite{MosleyPhD, Gerrits2015}.
Therefore, it is very valuable to explore such spectral information.
But in the past the spectral correlation content could not be observed, partially due to the long acquisition times required when measuring spectral correlations using two tunable bandpass filters \cite{Jin2013OE}.
Recently, a new technique for measuring spectral correlations in a Hong-Ou-Mandel (HOM) interference  \cite{Hong1987} was presented by Gerrits, et al \cite{Gerrits2015, Jin2015OE, Jin2016QST}.
This technique  has the merits of short measurement time and high spectral resolution, and therefore allows measuring and analyzing the spectral correlations from a HOM interference \cite{Gerrits2015, Jin2015OE, Jin2016QST}.
This technique also make it possible to observe the joint spectral information during the NOON-SI.

In this work, we apply the state-of-art spectral correlation measurement technique in the NOON-SI, and measure the joint spectral intensities (JSIs) at different positions of the interference.
We observe the interference pattern in the frequency domain clearly, while no interference appears in the time-domain measurement.
This implies that some new spectral information was indeed observed, although it was integrated and lost during the time-domain measurement.
%

%=============================================
\begin{figure*}[!tbp]
\centering
\includegraphics[width= 0.95\textwidth]{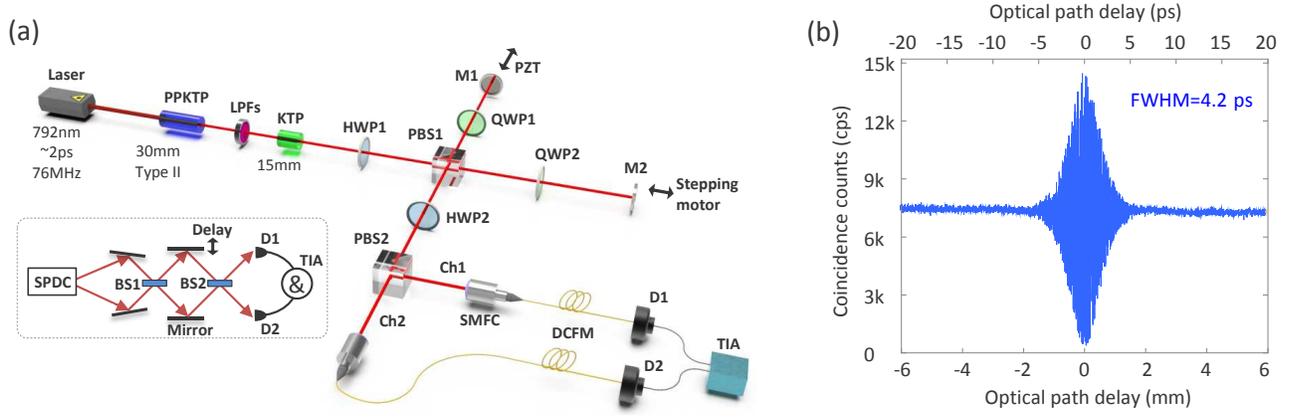}
\caption{(a): The experimental setup for NOON-SI using polarization-mode. LPF=long pass filter, HWP=half wave plate, M=mirror, QWP=quarter wave plate, PZT=piezo-electric linear actuator, PBS=polarization beam splitter, SMFC=single-mode fiber coupler,  DCFM=dispersion compensation fiber module, TIA=time interval analyzer.
The inset depicts a standard configuration of the NOON-SI using path-mode. BS=beam splitter.
(b): The NOON-SI pattern measured in time domain by scanning a stepping motor with a step length of 4 $\mu m$.
}
\label{experiment}
\end{figure*}
%=============================================
%

%
%
%=============================================
\begin{figure*}[tbp]
\centering
\includegraphics[width= 0.95\textwidth]{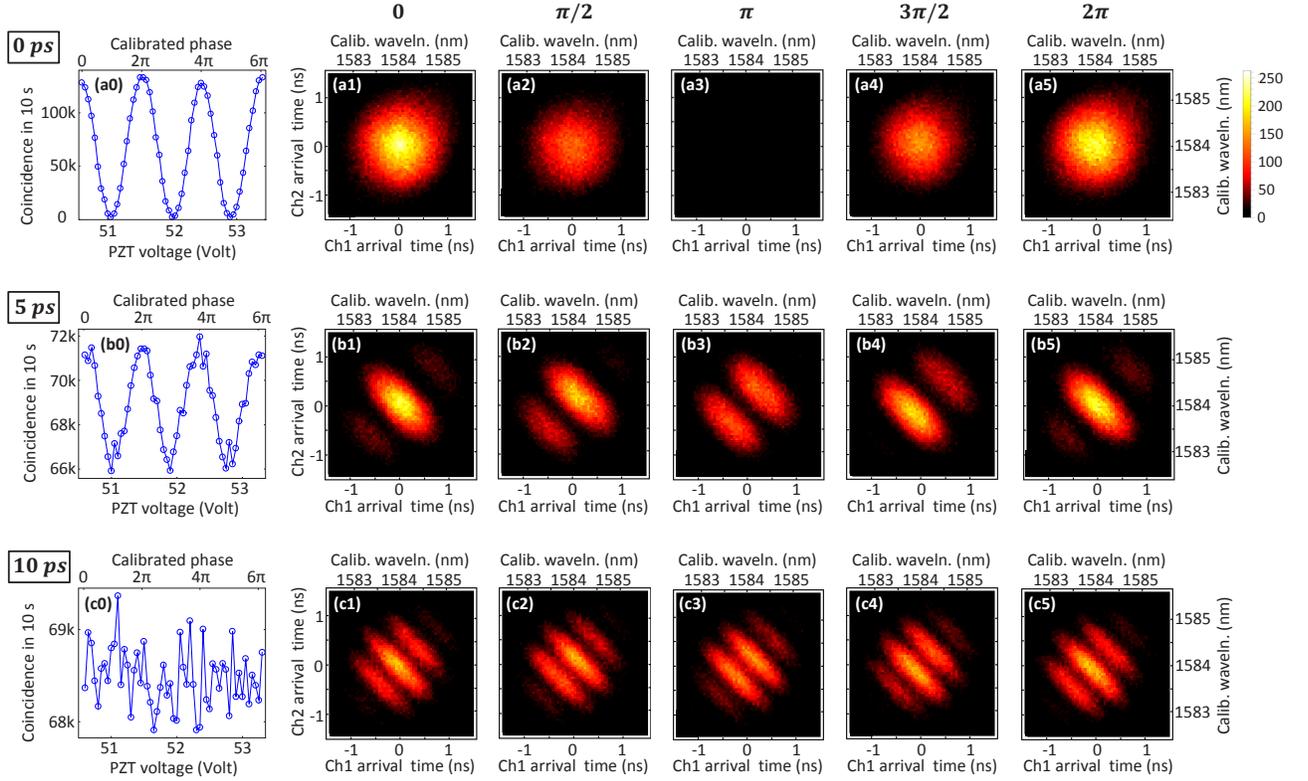}
\caption{  (a0, b0, c0): The experimental NOON-SI pattern measured in time domain by scanning a PZT with a step length several nm, when fix the stepping motor at the delay time of 0 ps, 5 ps and 10 ps, respectively. The visibility is 97.09 $\pm$ 0.03\%, 3.53$\pm$ 0.50\% and  0\%
(a1-a5, b1-b5, c1-c5): The measured JSIs at different phase delay from 0 to 2$\pi$  in (a0, b0, c0).
}
\label{results}
\end{figure*}
%=============================================

\textbf{\emph{Experiment and results}}
\noindent
The experimental setup for the NOON-SI is shown in Fig.\,\ref{experiment}(a).
Picosecond laser pulses (pulse repetition frequency = 76 MHz, wavelength = 792 nm, temporal duration $\sim$ 2 ps) from a mode-locked Titanium sapphire laser pump  a  30-mm-long PPKTP crystal with a poling period of 46.1 $\mu$m for type-II group-velocity-matched SPDC \cite{Konig2004,Jin2013OE}.
The down converted photons, i.e., the signal and idler photons, have orthogonal polarizations and degenerate wavelengths at 1584 nm.
To temporally overlap the signal and idler,  the downconverted biphotons pass through a 15-mm-long KTP crystal.
Then, the biphotons are mixed on a half wave plate (HWP1, at 22.5\,$^{\circ}$), which functions similarly as a beamsplitter in the NOON-SI using path-mode (BS1 in the inset of Fig.\,\ref{experiment}(a)).
After the biphotons are separated by a polarization beam splitter (PBS1), they are sent to a time delay system, which is composed of two quarter wave plates (QWP1 and QWP2, at 45\,$^{\circ}$) and two mirrors.
One of the mirrors (M1) is set on a PZT to achieve a scanning step  in the order  of nm, while the other one (M2) is on a stepping motor to realize a scanning step of $\mu$m.
Then, the biphotons are mixed again on the second half wave plate (HWP2, at 22.5\,$^{\circ}$), which functions as the BS2 in the inset.
Finally, the biphotons are separated by PBS2 and  collected into two channels of single-mode fibers (Ch1 and Ch2), which is connected to a dispersion spectrometer \cite{Avenhaus2009, Gerrits2011, Gerrits2015, Jin2015OE, Jin2016QST}, composed of a dispersion compensation fiber module (DCFM),  two superconducting nanowire single-photon detectors (SNSPDs) \cite{Miki2013} and a time interval analyzer (TIA).
The DCFM has a dispersion of 125.0 ps/km/nm (full dispersion of 941 ps/nm) and an insertion loss of 5.2 dB.
The SNSPDs have  detection efficiencies around 70\%  with dark counts less than 1 kHz.
Assuming a system jitter of 100 ps, the resolution of our fiber spectrometer is estimated as $0.11$ nm.
To keep the phase stability in the interference, we set the time delay system in an
enclosed paper box to avoid the air flow, such that the phase stability can be maintained for more than 5 minutes.

The PPKTP crystal is pumped with a power of 50 mW. With the insertion of DCFM, we obtain an average singles count rate of about 200 kcps, and a two-fold coincidence count rate of about 7 kcps.
First, we measure the NOON-SI in time domain, i.e., scan the time delay using a stepping motor  and record the  coincidence counts.
Fig.\,\ref{experiment}(b) shows the measured interference pattern with a full-width-at-half-maximum (FWHM) of 4.2 ps (1.3 mm).
Then, we characterize the NOON-SI in spectral domain, i.e., the JSI of the biphotons in Ch1 and Ch2 was measured at different delay time.
We focus on three delay time: 0 ps, 5 ps and 10 ps, by moving the stepping motor as the coarse adjustment and scan the PZT  as the fine adjustment.

Figure\,\ref{results}(a0) shows the interference pattern at 0 ps.
By driving the PZT with a voltage from 50 V to 53 V, the resultant relative phase delay is about 6$\pi$.
The interference visibility in Fig. \,\ref{results}(a0) is 97.09 $\pm$ 0.03\% and the oscillation period is 792 nm (0.462 fs), half of the biphoton's wavelength.
Then, we measure the JSI using the dispersion spectrometer.
Figure\,\ref{results}(a1-a5) shows the measured JSIs at the relative phase from 0 to 2$\pi$ with a step of $\pi$/2.
The horizontal (vertical) axis is the arrival time of the photons in Ch1 (Ch2).
The arrival time is ranging from -1.5 ns to 1.5 ns,  and the corresponding wavelength is from 1582 nm to 1586 nm.
Each figure was accumulated for 60 seconds.
 The JSI at 0 ($2\pi$) is the brightest and at $\pi$ is the darkest.
The JSI in Fig. \,\ref{results}(a1) is equal to the JSI of the biphoton source.

Figure\,\ref{results}(b0) shows the measured coincidence counts as a function of time delay at 5 ps.
The visibility is 3.53 $\pm$ 0.50\%, which can also be confirmed in Fig.\,\ref{experiment}(b).
Figure\,\ref{results}(b1-b5) shows the measured JSIs at the phase of 0, $\frac{\pi}{2}$, $\pi$, $\frac{3\pi}{2}$, and $2\pi$  in Fig.\,\ref{results}(b0).
It can be observed clearly that the distribution of the JSIs  is splitting along the diagonal axis. The JSI is phase sensitive, i.e, the distribution in the JSI is different for different phases.
At  $\frac{\pi}{2}$, the up-right section is brighter, while at $\frac{3\pi}{2}$, the down-left section is brighter.

Figure\,\ref{results}(c0) shows the temporal interference pattern at the time delay of 10 ps.
The visibility is almost zero in Fig.\,\ref{results}(c0), as also confirmed in Fig.\,\ref{experiment}(b).
Figure\,\ref{results}(c1-c5) shows the measured JSIs at the phase of 0, $\frac{\pi}{2}$, $\pi$, $\frac{3\pi}{2}$, and $2\pi$ in Fig.\,\ref{results}(c0).
It can be noticed that the JSI is separated into odd-number sections  at 0 ($2\pi$), and even-number sections at $\pi$.
It can be concluded that when the time-domain interference visibility is almost zero, the interference patterns in the spectral domain are still very clear.

\textbf{\emph{Theoretical analysis}}
\noindent
The experimental results in Fig.\,\ref{results} is well explained in theory.
Assume the signal and idler photons from the SPDC has a joint spectral amplitude (JSA) of $f(\omega _s ,\omega _i )$, where $\omega$ is the angular frequency; the subscripts $s$ and $i$ denote the signal and idler. For simplicity, we further assume the JSA has the exchanging symmetry of $f(\omega _s ,\omega _i )=f(\omega _i ,\omega _s )$. After a long calculation using multi-spectral-mode theory \cite{Jin2018Optica, Ou2007},
%,
the two-fold coincidence probability $P$ as a time delay $\tau$ in a NOON-SI can be described by
\begin{equation}\label{eq1}
P(\tau ) = \frac{1}{2}\int_0^\infty  \int_0^\infty  d\omega _s  d\omega _i \left|f(\omega _s ,\omega _i )\right|^2 [1+ \cos(\omega _s  +\omega _i )\tau  ].
\end{equation}
The JSI $I(\omega _s ,\omega _i, \tau)$ during the NOON-SI can be written as
\begin{equation}\label{eq2}
I(\omega _s, \omega _i, \tau ) =  \frac{1}{2}  \left|  f(\omega _s ,\omega _i )\right|^2 [1+\cos(\omega _s  +\omega _i )\tau  ].
\end{equation}
Using the parameters of the pump laser and the PPKTP crystal, we can calculate the JSA.
With the JSA, the time-domain interference patterns in Fig.\,\ref{experiment}(b) and Fig.\,\ref{results}(a0, b0, c0)  can be simulated using Eq.(\ref{eq1}). The frequency-domain interference patterns in Fig.\,\ref{results}(a1-a5, b1-b5, c1-c5)  can be simulated by using Eq.(\ref{eq2}).

\begin{figure*}[tbp]
\centering
\includegraphics[width= 0.95\textwidth]{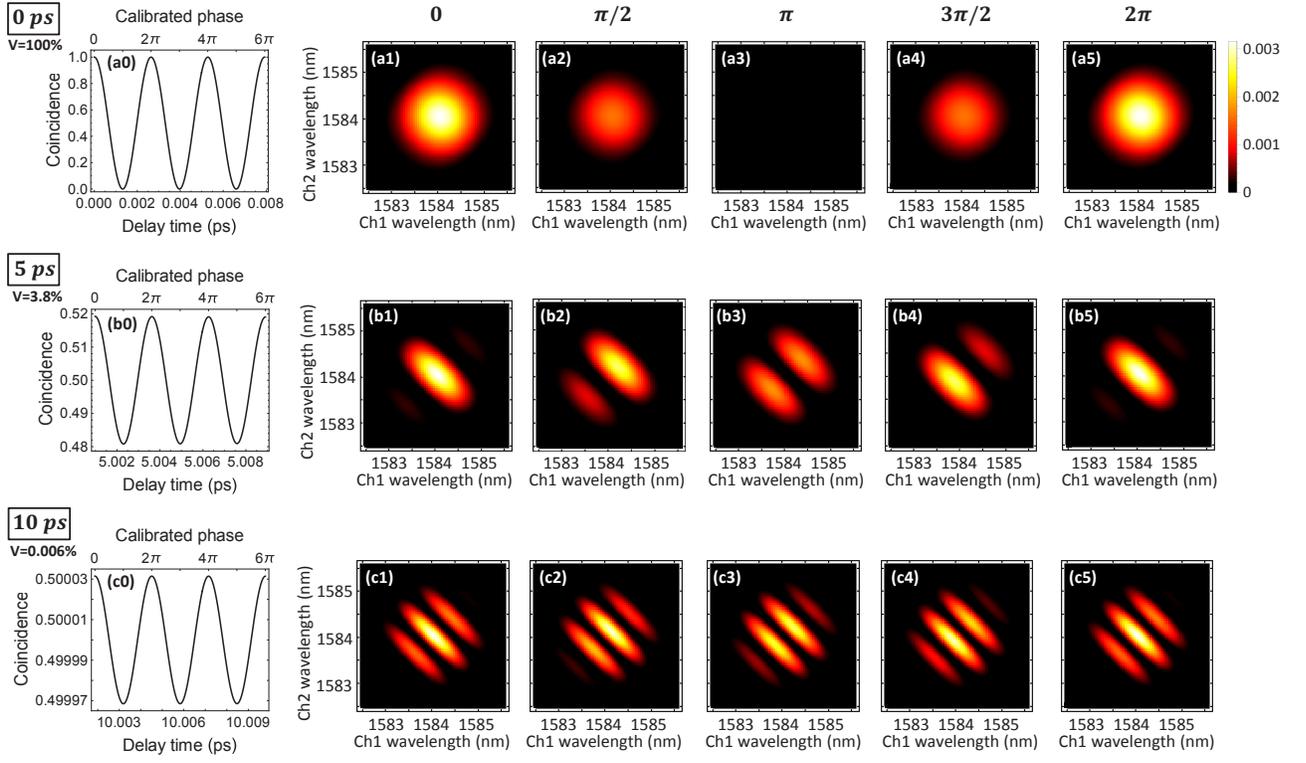}
\caption{  (a0, b0, c0): The theoretically simulated NOON-SI pattern in time domain at the delay time of 0 ps, 5 ps and 10 ps, respectively. The visibilities are 100\%, 3.8\% and  0.006\%.
(a1-a5, b1-b5, c1-c5): The simulated JSIs at different phase delay from 0 to 2$\pi$  in (a0, b0, c0).
}
\label{simulation}
\end{figure*}

Figure\,\ref{simulation} shows the simulated results.
In Fig.\,\ref{simulation}(a0, b0, c0), the visibility are 100\%, 3.8\% and  0.006\%, which correspond well with the experimental results in Fig.\,\ref{results}(a0, b0, c0).
Figure\,\ref{simulation}(a1-a5, b1-b5, c1-c5) shows the simulated JSIs at different phase delay from 0 to 2$\pi$  in (a0, b0, c0).
The measured JSIs are consistent with the simulated ones.
The experimentally measured JSIs are ``fatter" than the simulated JSIs, since the resolution of the fiber spectral meter is limited.

%
%
%=============================================
%=============================================

\textbf{\emph{Discussion}}
\noindent
We compare the traditional {\it non-spectrally resolved (NSR)} NOON-SI and  the {\it spectrally resolved} NOON-SI in our scheme by analyzing the Fisher information, which has been widely used to evaluate the phase sensing ability of the NOON-SI \cite{Xiang2013}.
The
Fisher information for the {\it non-spectrally resolved} NOON-SI ($F_{NSR}$) can be calculated as:
\begin{equation}
F_{NSR} (\tau ) = \frac{{\left[ {P'_{NSR} (\tau )} \right]^2 }}{{P_{NSR} (\tau )[1 - P_{NSR} (\tau )]}},
\end{equation}
where $P_{NSR}(\tau ) = P(\tau )$ and $P'_{NSR}$ is the differentiation of $P_{NSR}$ with respect to the time $\tau$. %See Appendix for more details.
Fisher information for the {\it spectrally resolved} NOON-SI ($F_{SR}$) is
\begin{equation}
F_{SR} (\tau ) = \int_0^\infty  \int_0^\infty |f(\omega _s ,\omega _i )|^2 \times (\omega _s {\rm{ + }}\omega _i )^2 d\omega _s d\omega _i.
\end{equation}
With these two equations and the theoretical $f(\omega_s, \omega_i)$, we simulated $F_{NSR}$   and $F_{SR}$ .
It is noteworthy that  $F_{SR}$ equals to the maximal value of $F_{NSR}$ and does not degrade in delay. This feature implies that the spectrally resolved techniques enable the acquisition of higher Fisher information in wider range of delay time.
This advantage is very useful for time or phase estimation in wide range.

It is interesting to compare this spectrally resolved NOON-SI with the result in the spectrally resolved HOM interference in Refs. \cite{Gerrits2015, Jin2015OE, Jin2016QST}. In the NOON-SI, the JSI is splitting along the anti-diagonal direction, while the JSI is splitting along the diagonal direction in the HOM interference. This difference is result from the facts that the NOON-SI is a sum-frequency interference, which is described in Eq.(\ref{eq1}), while the HOMI is a difference-frequency interference, as explained in Refs. \cite{Gerrits2015, Jin2015OE, Jin2016QST, Jin2018Optica, Giovannetti2002}.
It can be observed that the distribution of the JSI is changing at different phase delay. This phase dependent feature is totally different from the case of HOM interference \cite{Gerrits2015, Jin2015OE, Jin2016QST}, which is phase independent.
The spectrally resolved HOM interference has been applied for the situation in which the phase insensitivity is required, e.g., the generation and distribution of frequency-entangled qudits \cite{Jin2016QST}, efficient  quantum-optical coherence tomography \cite{Yepiz-Graciano2020PRJ}.

We noticed the recent report on the spectrally-resolved quantum white-light interferometry (WLI) for optical chromatic dispersion measurement, which has a better precision than the classical WLI and can be used for absolute parameter determination \cite{Kaiser2018light}. In Ref. \cite{Kaiser2018light} only one spectrometer was utilized, i.e., one-dimensional spectrally-resolved. In contrast, our measurement are two-dimensional spectrally resolved, where two spectrometer are utilized to measure the spectral correlation between the signal and idler photons. This two dimensional spectral resolving technique may also has the potential for higher precision %\textcolor{red}{absolute}
parameter estimation in wider range in the future.

It is important to discuss the feasibility of the future applications.
The interference visibility is almost 0 in time domain at 10 ps in Fig.\,\ref{results}(c0),  however, the visibility in spectral domain in Fig.\,\ref{results}(c1-c5) is still very high.
This implies that the spectral measurement can exploit some new information that has never been measured in the time domain.
This extra spectral information might be an important supplementary to the time-domain information, and might be used for promising applications, such as remote synchronization \cite{Quan2016},  quantum spectroscopy \cite{Dinani2016, Whittaker2017, Whittaker2017} and quantum metrology \cite{Dowling2008, Taylor2016} in the future. % Misiaszek2018

\textbf{\emph{Conclusion}}
\noindent
We have experimentally and theoretically demonstrated  spectrally resolved NOON-SI with a spontaneous parametric down conversion source at 1584 nm wavelength.   It was confirmed that, while there is no interference in time domain, the interference visibility in spectral domain is still very high.  The spectrally resolved NOON-SI has a higher Fisher information than the non-spectrally resolved NOON-SI at the position far from the interference center, therefore has a better performance for
time or phase estimation in a wider range.  This spectrally resolved NOON-SI scheme may be applied for remote synchronization, absolute parameter estimation, quantum spectroscopy,  and quantum metrology.

\textbf{\emph{Acknowledgements}}
The authors thank Thomas Gerrits  and  Kentaro Wakui for helpful discussions.
This work is  partially supported by the ImPACT Program of Council for Science, Technology and Innovation (Cabinet Office, Government of Japan),
by the National Key R$\&$D Program of China (Grant No. 2018YFA0307400), the National Natural Science Foundations of China (Grant Nos. 91836102, 11704290, 12074299, 61775025, 12074058), and by a fund from the Educational Department of Hubei Province, China (Grant No. D20161504).

\end{document}